\documentclass[english]{article}
\usepackage[T1]{fontenc}
\usepackage[latin1]{inputenc}
\usepackage{graphicx}

\makeatletter

\providecommand{\tabularnewline}{\\}

\makeatletter

\makeatletter

\makeatletter
\usepackage{geometry}

\makeatother

\makeatother

\makeatother

\usepackage{babel}
\makeatother
\begin{document}

\title{\textbf{Effects of partial triple excitations in atomic coupled cluster
calculations}}

\author{Chiranjib Sur\\
 \emph{Department of Astronomy, Ohio State University, Columbus, Ohio,
43210, USA}\\
 \\
 Rajat K Chaudhuri\\
 \emph{Indian Institute of Astrophysics, Koramangala, Bangalore, 560
034, India}}

\date{Ref : Chem. Phys. Lett., vol 442, 150 (2007) }

\maketitle
\begin{abstract}
In this article we study the effects of higher body excitations in
the relativistic CC calculations for atoms and ions with one valence
electron using Fock-space CCSD, CCSD(T) and its unitary variants.
The present study demonstrates that CCSD(T) estimates the ionization
potentials (IPs) and the valence electron removal energies quite accurately
for alkali atoms and singly ionized alkaline earth ions, but yields
unphysical energy levels for atoms and/or ions with partially filled
sub-shell like C II. We further demonstrate that the higher body excitation
effects can be incorporated more effectively through the unitary coupled
cluster theory (UCC) compared to the CCSD(T) method.

\textbf{PACS numbers} : 31.15.Ar, 31.15.Dv 
\end{abstract}

\section{\label{intro}Introduction}

The relativistic coupled cluster (CC) method has emerged as one of
the most powerful and effective tool for accurate treatment of electron
correlation and relativistic effects in many-electron systems \cite{kaldor-book}.
The CC is an all-order non-perturbative scheme, and therefore, the
higher order electron correlation effects can be incorporated more
efficiently than using the order-by-order diagrammatic many-body perturbation
theory (MBPT) \cite{lindgren-book}. The method is also size-extensive
\cite{size-extensive}, a property which has been found to be crucial
for accurate determination of state energies of atoms and related
spectroscopic constants. The incorporation of the singly and doubly
excited cluster operators (SD) within the single reference (SR) CC
framework provides a reasonably accurate and reliable description
of the electron correlation for non-degenerate states, and is one
of the most extensively used class of CC approaches.

The CCSD scheme often fails to provide results of sufficient accuracy
and even breaks down for highly correlated systems \cite{break-down}.
In recent years, considerable progress has been made in pushing the
boundaries of its applicability through the inclusion of higher order
clusters in CC methodology both in the singles \cite{srcc} as well
as in the multi-reference formulations \cite{lindgren,mukherjee,mrcc}.
In this regard, the non-iterative approaches like CCSD(T) \cite{ccsd(t)}
offer substantial time savings compared to their full CC counterparts,
namely, CCSDT (CC with singles, doubles and triples) \cite{ccsdt}
and CCSDTQ (CC with singles, doubles, triples and quadruples) \cite{ccsdtq}.
In this letter, we demonstrate that a unitary form of the wave operator
for the closed shell component of the CC-ansatz can incorporate the
effects of partial triples more efficiently than the non-iterative
perturbative connected triples corrections, CCSD(T).

To illustrate our findings we compute the ionization potentials (IPs)
and valence electron removal energies of C II and Rb I using the Fock-space
multi-reference coupled cluster (FSMRCC) method for one electron attachment
process (M$^{+n}$+e$\rightarrow$M$^{+(n-1)}+\Delta E$). The ground
state electronic configurations of these two systems reveal that C
II has an unfilled L-shell whereas Rb I has a completely filled N-shell
followed by one electron ($5s$) in the O shell. In this article we
demonstrate that partial triple excitation affects the determination
of the ionization potential (IP), the valence electron removal energies
and therefore the excitation energies (EEs) of the excited states
depending on the vacancies in the principal shell. The behavior of
partial triple excitations through CCSD(T) and UCCSD (UCC with singles
and doubles excitations) for these kind of systems is also addressed.
To our knowledge, this work is the first attempt to establish the
effects of partial triple excitations through the core and valence
excitations for determining the state energies for single valence
atoms having a filled or unfilled sub-shell.

The structure of this article is as follows : Section \ref{OSCC}
provides a brief outline of the Fock-space CC theory followed by the
higher body effects in section \ref{Higher-excit}. Subsections \ref{unitary}
and \ref{partial-trip} presents the unitary coupled cluster theory
and the results are discussed in the subsequent section.

\section{\label{OSCC}\label{OSCC}Fock-space multi-reference coupled cluster
theory for one-electron attachment process}

\noindent Relativistic extension of coupled cluster (CC) theory is
based on the no-virtual-pair approximation (NVPA) along with appropriate
modification of orbital form and potential terms \cite{eliav}. Relativistic
CC theory begins with Dirac-Coulomb Hamiltonian (H) for an $N$ electron
atom which is expressed as

\begin{equation}
{\mbox{H}}=\sum_{i=1}^{N}\left[c\vec{\alpha_{i}}\cdot\vec{p}_{i}+\beta mc^{2}+V_{\mathrm{Nuc}}(r_{i})\right]+\sum_{i<j}^{N}\frac{e^{2}}{r_{ij}}\label{eq1}\end{equation}
 with all the standard notations often used. The normal ordered form
of the above Hamiltonian is given by \begin{equation}
{\mathcal{H}}={\mbox{H}}-\langle\Phi|{\mbox{H}}|\Phi\rangle={\mbox{H}}-{\mbox{E}}_{\tiny\mbox{DF}}=\sum_{ij}\langle i|f|j\rangle\left\{ a_{i}^{\dagger}a_{j}\right\} +\frac{1}{4}\sum_{i,j,k,l}\langle ij||kl\rangle\left\{ a_{i}^{\dagger}a_{j}^{\dagger}a_{l}a_{k}\right\} .\label{eq2}\end{equation}
 where \begin{equation}
\langle ij||kl\rangle=\langle ij|\frac{1}{r_{12}}|kl\rangle-\langle ij|\frac{1}{r_{12}}|lk\rangle.\label{eq3}\end{equation}
 Here E$_{\tiny\mbox{DF}}$ is the Dirac-Fock energy, $f$ is the
one-electron Fock operator, $a_{i}(a_{i}^{\dagger})$ is the annihilation
(creation) operator (with respect to the Dirac-Fock state as the vacuum)
for the $i$th electron and $\left\{ \cdots\right\} $ denotes the
normal ordering of the creation/annihilation operators.

\noindent Since the FSMRCC theory has been described elsewhere \cite{lindgren,mukherjee,Haque},
we provide a brief review of this method. The FSMRCC theory is based
on the concept of common vacuum for both the N and N$\pm$m electron
systems, which allows us to formulate a direct method for energy differences.
In this method the holes and particles are defined with respect to
the common vacuum for both the N and N$\pm$m electron systems. Model
space of a (m,n) Fock-space contains determinants with $m$ holes
and $n$ particles distributed within a set of what are termed as
{\em active} orbitals. For example, in this present article, we
are dealing with (0,1) Fock-space which is a complete model space
(CMS) by construction and is given by

\begin{equation}
|\Psi_{\mu}^{(0,1)}\rangle=\sum_{i}{\mbox{C}}_{i\mu}|\Phi_{i}^{(0,1)}\rangle\label{eq4}\end{equation}
 where ${\mbox{C}}_{i\mu}$'s are the coefficients of $\Psi_{\mu}^{(0,1)}$
and $\Phi_{i}^{(0,1)}$'s are the model space configurations. The
dynamical electron correlation effects are introduced through the
{\em valence-universal} wave-operator $\Omega$ \cite{lindgren,mukherjee}

\noindent \begin{equation}
\Omega={\{\exp({\tilde{\mbox{S}}})}\}\label{eq5}\end{equation}
 where \begin{equation}
{\tilde{\mbox{S}}}=\sum_{k=0}^{m}\sum_{l=0}^{n}\mbox{S}^{(k,l)}={\mbox{S}}^{(0,0)}+{\mbox{S}}^{(0,1)}+{\mbox{S}}^{(1,0)}+\cdots\label{eq6}\end{equation}
 At this juncture, it is convenient to single out the core-cluster
amplitudes {\mbox{S}$^{(0,0)}$} and call them T. The rest of
the cluster amplitudes will henceforth be called S. Since $\Omega$
is in normal order, we can rewrite Eq.(\ref{eq5}) as \begin{equation}
\Omega=\mbox{exp(T)}{\{\mbox{exp}({\mbox{S}})}\}\label{eq7}\end{equation}
 The {}``valence-universal'' wave-operator $\Omega$ in Eq.(\ref{eq7})
is parametrized in such a way that the states generated by its action
on the reference space satisfy the Fock-space Bloch equation \begin{equation}
{\mbox{H}}\Omega{\mbox{P}}^{\mbox{(k,l)}}=\Omega{\mbox{P}}^{\mbox{(k,l)}}{\mbox{H}}_{\tiny\mbox{eff}}{\mbox{P}}^{\mbox{(k,l)}}\label{eq8}\end{equation}
 where \begin{equation}
{\mbox{H}}_{\tiny\mbox{eff}}={\mbox{P}}^{\mbox{(k,l)}}{\mbox{H}}\Omega{\mbox{P}}^{\mbox{(k,l)}}.\label{eq9}\end{equation}
 Eq.(\ref{eq8}) is valid for all (k,l) starting from k=l=0, the {\em
core} problem to some desired {\em parent} model space, with k=m,
l=n, say. In this present calculation, we truncate Eq.(\ref{eq6})
at $m=0$ and $n=1$. The operator P$^{\mbox{(k,l)}}$ in Eqs. (\ref{eq8})
and (\ref{eq9}) is the model space projector for k-hole and l-particle
model space which satisfies

\noindent \begin{equation}
{\mbox{P}}^{\mbox{(k,l)}}\Omega{\mbox{P}}^{\mbox{(k,l)}}={\mbox{P}}^{\mbox{(k,l)}}.\label{eq10}\end{equation}
 To formulate the theory for direct energy differences, we pre-multiply
Eq.(\ref{eq8}) by exp(-T) (i.e., $\Omega_{c}^{-1}$) and get \begin{equation}
{\overline{\mbox{H}}}\Omega_{v}{\mbox{P}}^{\mbox{(k,l)}}=\Omega_{v}{\mbox{P}}^{\mbox{(k,l)}}{\mbox{H}}_{\tiny\mbox{eff}}{\mbox{P}}^{\mbox{(k,l)}}\hspace{0.2in}\forall(k,l)\ne(0,0)\label{eq11}\end{equation}
 where ${\overline{\mbox{H}}}$=e$^{\mbox{-T}}$ H e$^{\mbox{T}}$.
Since ${\overline{\mbox{H}}}$ can be partitioned into a connected
operator ${\tilde{\mbox{H}}}$ and E$_{\tiny\mbox{ref/gr}}$ (N-electron
closed-shell reference or ground state energy), we likewise define
${\tilde{\mbox{H}}}_{\tiny\mbox{eff}}$ as \begin{equation}
{\mbox{H}}_{\tiny\mbox{eff}}={\tilde{\mbox{H}}}_{\tiny\mbox{eff}}+{\mbox{E}}_{\tiny\mbox{ref/gr}}.\label{eq12}\end{equation}
 Substituting Eq.(\ref{eq12}) in Eq.(\ref{eq11}) we obtain the Fock-space
Bloch equation for energy differences: \begin{equation}
{\tilde{\mbox{H}}}\Omega_{v}{\mbox{P}}^{\mbox{(k,l)}}=\Omega_{v}{\mbox{P}}^{\mbox{(k,l)}}{\tilde{\mbox{H}}}_{\tiny\mbox{eff}}{\mbox{P}}^{\mbox{(k,l)}}.\label{eq13}\end{equation}
 Eqs. (\ref{eq8}) and (\ref{eq13}) are solved by Bloch projection
method, involving the left projection of the equation with P$^{\mbox{(k,l)}}$
and its orthogonal complement Q$^{\mbox{(k,l)}}$ to obtain the effective
Hamiltonian and the cluster amplitudes, respectively. At this juncture,
we recall that the cluster amplitudes in FSMRCC are generated hierarchically
through the \emph{subsystem embedding condition} (SEC) \cite{Haque,SEC}
which is equivalent to the  \emph{valence universality} condition
used by Lindgren\cite{lindgren} in his formulation. For example,
in the present application, we first solve the Fock-space CC for k=l=0
to obtain the core-cluster amplitudes T. The operator ${\tilde{\mbox{H}}}$
and ${\tilde{\mbox{H}}}_{\tiny\mbox{eff}}$ are then constructed from
this core-cluster amplitudes T to solve the Eq. (\ref{eq13}) for
k=0, l=1 to determine S$^{(0,1)}$ amplitudes. The effective Hamiltonian
constructed from H, T, and S$^{(0,1)}$ is then diagonalized within
the model space to obtained the desired eigenvalues and eigenvectors
\begin{equation}
{\tilde{\mbox{H}}}_{\tiny\mbox{eff}}{\mbox{C}}^{(0,1)}={\mbox{C}}^{(0,1)}{\mbox{E}}.\label{eq14}\end{equation}

\section{\label{Higher-excit}Higher order excitations}

\noindent It is now widely recognized that the effects of higher body
clusters must be included in CC calculations to improve the accuracy
of the predicted/computed quantities. Here by the term `higher body
effects', we mean effects from triple, quadruple excitations etc.
In this letter, we shall restrict ourselves only to triple excitations
for the time being and will comment on other higher excitations later.
The most straightforward approach is to include the full three body
excitation operators T$_{3}$ and S$_{3}$ in the CC ansatz via T=T$_{1}$+T$_{2}$+T$_{3}$
and S=S$_{1}$+S$_{2}$+S$_{3}$. This direct approach, known as CCSDT,
is computationally very expensive.

In this article we have used the unitary ansatz to simulate the effects
of triples and some other higher body excitations, \emph{e.g.}, quadruples
etc. in the core sector. In addition, we have also considered the
effects of partial triple excitations in a perturbative way for the
(0,1) valence sector known as CCSD(T). These are discussed in the
next two subsections.

\subsection{\label{unitary}Higher body excitations through unitary ansatz}

Unitary coupled-cluster (UCC) theory was first proposed by Kutzelnigg
\cite{kutzelnigg-ucc}. In this theory, the effective Hamiltonian
is Hermitian by construction and the energy which is the expectation
value of this operator in the reference state is an upper bound to
the ground state energy \cite{crawford}.

The normal ordered dressed Hamiltonian is expressed by the Baker-Hausdorff-Campbell
expansion in CC theory as \begin{eqnarray}
{\overline{\mathcal{H}}} &  & =e^{\mbox{-T}}{\mathcal{H}}e^{\mbox{T}}\nonumber \\
 &  & ={\mathcal{H}}+[{\mathcal{H}},{\mbox{T}}]+\frac{1}{2!}[[{\mathcal{H}},{\mbox{T}}],{\mbox{T}}]\nonumber \\
 &  & +\frac{1}{3!}[[[{\mathcal{H}},{\mbox{T}}],{\mbox{T}}],{\mbox{T}}]+\frac{1}{4!}[[[[{\mathcal{H}},{\mbox{T}}],{\mbox{T}}],{\mbox{T}}],{\mbox{T}}].\end{eqnarray}
 In UCC, the operator T is replaced by $\sigma_{c}={\mbox{T}}-{\mbox{T}}^{\dagger}$
in the above equation. As a result, ${\overline{\mathcal{H}}}$ is
expressed in terms of a non-terminating series of commutators. For
practical reasons, one truncates the series after some finite order.
Truncation at the \emph{n}-th order commutator leads to the nomenclature
UCC(\emph{n}).

\begin{figure}[ht]
\begin{centering}
\includegraphics[width=8.5cm]{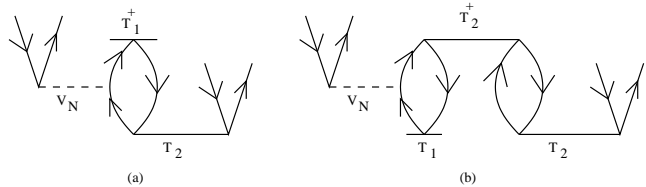} 
\par\end{centering}

\caption{\label{fig-triples}Typical effective triples and quadruples diagrams
arising from UCC(\emph{3}). $\mathrm{V_{N}}$ represents the Coulomb
vertex.}
\end{figure}

Using UCC(\emph{3}) approximation and without modifying the last term
of the above expression, one can show that the dressed Hamiltonian
takes the form \begin{equation}
{\overline{\mathcal{H}}={\mathcal{H}}+\overline{{\mathcal{H}}{\mbox{T}}}+\frac{1}{2!}(\overline{\overline{{\mathcal{H}}{\mbox{T}}}{\mbox{T}}}+2\overline{\overline{{\mbox{T}}^{\dagger}{\mathcal{H}}}{\mbox{T}}})+\frac{1}{3!}(\overline{\overline{\overline{{\mathcal{H}}{\mbox{T}}}{\mbox{T}}}{\mbox{T}}}+3\overline{{\mbox{T}}^{\dagger}\overline{{\mbox{T}}^{\dagger}\overline{{\mathcal{H}}{\mbox{T}}}}}+3\overline{{\mbox{T}}^{\dagger}\overline{\overline{\mathcal{H}}{\mbox{T}}}{\mbox{T}}}})+\frac{1}{4!}\overline{\overline{\overline{\overline{{\mathcal{H}}{\mbox{T}}}{\mbox{T}}}{\mbox{T}}}{\mbox{T}}}\label{eq16}\end{equation}
 Here `overline' denotes the contraction between two sets of operators.
For example, the term $\overline{{\mathcal{H}}{\mbox{T}}}$ corresponds
to the contraction between the operators ${\mathcal{H}}$ and T. A
typical contribution to the term $\overline{\overline{{\mathcal{H}}{\mbox{T}_{2}}}{\mbox{T}_{2}}}$
is given by

\begin{equation}
B_{ab}^{pq}=\frac{1}{2}\sum_{dgrs}V_{dgrs}t_{ad}^{pr}t_{gb}^{sq}.\label{eq17}\end{equation}
 Here $V_{dgrs}$ is the two-electron Coulomb integral and $t_{ad}^{pr}$
is the cluster amplitude corresponding to a simultaneous excitation
of two electrons from orbital $a\rightarrow p$ and $d\rightarrow r$,
respectively. This term is common both to CCSD and UCCSD whereas the
latter contains some higher order terms containing T$^{\dagger}$
which are not present in the CCSD expansion of ${\overline{\mathcal{H}}}$
\cite{csur-ucc}. Diagrammatic techniques are used to obtain all the
terms which contribute to this specific contribution. Fig. \ref{fig-triples}
shows two typical diagrams arises from UCC(3) which correspond to
a subset of effective triple (\ref{fig-triples}a) and quadruple excitation
(\ref{fig-triples}b) effects respectively.

\subsection{\label{partial-trip}Higher order excitations in the valence sector}

\begin{figure}
\begin{centering}
\includegraphics[scale=0.5]{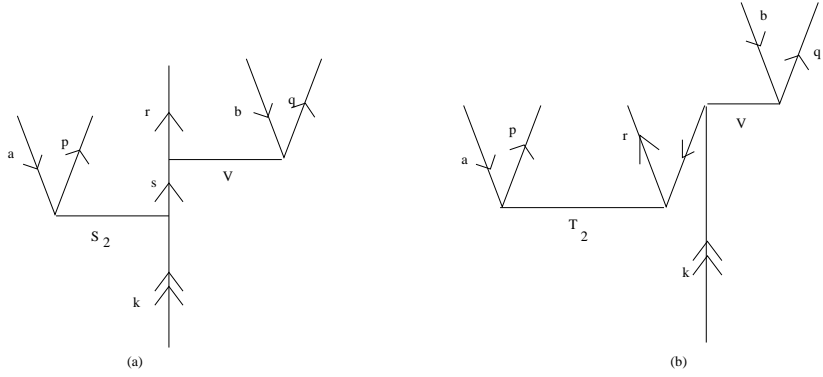} 
\par\end{centering}

\caption{\label{diag-triples}Some typical important diagrams which arise
due to the inclusion of triple excitations through Eq.(\ref{eq21}).
In this diagram $\mathrm{V}$ denotes the Coulomb vertex.}
\end{figure}

Triple excitations are included in the open shell CC amplitude which
correspond to the correlation to the valence orbitals, by an approximation
that is similar in spirit to CCSD(T) \cite{ccsd(t)}. The approximate
valence triple excitation amplitude is given by

\begin{equation}
{\mbox{S}}_{abk}^{pqr}=\frac{{\{{\overline{V{\mbox{T}}_{2}}}\}_{abk}^{pqr}}+{\{{\overline{V{\mbox{S}}_{2}}}}\}_{abk}^{pqr}}{\varepsilon_{a}+\varepsilon_{b}+\varepsilon_{k}-\varepsilon_{p}-\varepsilon_{q}-\varepsilon_{r}},\label{eq21}\end{equation}
 where S$_{abk}^{pqr}$ are the amplitudes corresponding to the simultaneous
excitation of orbitals $a,b,k$ to $p,q,r$, respectively; $\overline{V{\mbox{T}}_{2}}$
and $\overline{V{\mbox{S}}_{2}}$ are the correlated composites involving
$V$ and T, and $V$ and S respectively where $V$ is the two electron
Coulomb integral and $\varepsilon$'s are the orbital energies. The
above amplitudes (some representative diagrams are given in Fig. \ref{diag-triples})
are added appropriately to the singles and doubles S amplitude determining
equations and these equations are then solved iteratively.

\section{\label{results}Results - CC calculations for atoms with single valence
electron}

In this article we have considered two systems C II and Rb I. C II
is the singly ionized C atom and the ground state has an atomic structure
like Boron (B I) : $1s^{2}2s^{2}2p_{1/2}$ where as the electronic
structure of Rb I ground state is $[\mathrm{Kr]}5s$ \emph{i.e.} $1s^{2}2s^{2}2p^{6}3s^{2}3p^{6}3d^{10}4s^{2}4p^{6}5s$.
As because the ground state of C II and Rb I is an open-shell doublet,
we begin with C III and Rb II which defines the (0h,0p) valence sector.
We then employ the open-shell Fock space CC theory for one electron
attachment process to compute the ionization potentials (IPs) of the
ground state and excitation energies (EEs) of the the first few excited
states of C II and Rb I, which are given in tables \ref{CII-results}
and \ref{RbI-results} respectively. We have also calculated those
quantities taking into account the effect of partial triple excitations
for the valence electron and are listed as CCSD(T)/UCCSD(T) in the
corresponding tables.

The Dirac-Fock equations are first solved for closed shell ions (C
III and Rb II), which defines the (0,0) sector of the Fock space.
The ion is then correlated using the closed shell CCSD/UCCSD, after
which one-electron is added following the Fock-space strategy:

\begin{equation}
\mathrm{M^{+n}(0,0)}+e\longrightarrow\mathrm{M^{+(n-1)}(0,1)}.\end{equation}

Both the DF and relativistic CC programs utilize the angular momentum
decomposition of the wave-functions and CC equations. Using the Jucys-
Levinson-Vanagas (JLV) theorem \cite{jlv}, the Goldstone diagrams
are expressed as a products of angular momentum diagrams and reduced
matrix element. This procedure simplifies the computational complexity
of the DF and CC equations. We use the kinetic balance condition to
avoid the {}``variational collapse'' \cite{kineticbalance-lee}.

\begin{table}

\caption{\label{basis}Total number of the basis functions and the even tempering
parameters ($\alpha_{0}$ and $\beta$ ) used in the calculations.
GTOs stand for the Gaussian type orbitals used to generate the DF
wave-functions. `Active orbitals' refer to the number of orbitals
used in the CC/UCC calculations. The parameters $\alpha_{0}$ and
$\beta$ for C II (Rb I) are 0.005 (0.00523) and 2.25 (2.09) respectively
which are used in Eq. (\ref{eq21}) to generate the DF orbitals. }

\begin{centering}
\begin{tabular}{llllllllll}
\hline 
&
$s_{1/2}$ &
$p_{1/2}$&
$p_{3/2}$&
$d_{3/2}$&
$d_{5/2}$&
$f_{5/2}$&
$f_{7/2}$&
$g_{7/2}$&
$g_{9/2}$\tabularnewline
\hline
\hline 
C II&
&
&
&
&
&
&
&
&
\tabularnewline
No. of GTOs&
35&
32&
32&
25&
25&
25&
25&
20&
20\tabularnewline
Active orbitals&
14&
13&
13&
11&
11&
9&
9&
6&
6\tabularnewline
Rb I&
&
&
&
&
&
&
&
&
\tabularnewline
No. of GTOs&
38&
35&
35&
25&
25&
25&
25&
20&
20\tabularnewline
Active orbitals&
14&
12&
12&
10&
10&
9&
9&
6&
6\tabularnewline
\hline
\hline 
&
&
&
&
&
&
&
&
&
\tabularnewline
\end{tabular}
\par\end{centering}
\end{table}

In the actual computation, the DF ground state and excited state properties
are computed using the finite basis set expansion method (FBSE) \cite{rajat-gauss}
with a large basis set of Gaussian type functions (GFs) of the form
\begin{equation}
F_{i,k}(r)=r^{k}\cdot e^{-\alpha_{i}r^{2}}\label{eq20}\end{equation}
 with $k=0,1,\dots$ for $s,p,\dots$ type functions, respectively.
For the exponents, the even tempering condition \begin{equation}
\alpha_{i}=\alpha_{0}\beta^{i-1}\label{eq21}\end{equation}
 is applied. The nucleus has a finite structure and is described by
the two parameter Fermi nuclear distribution

\begin{equation}
\rho=\frac{\rho_{0}}{1+\exp((r-c)/a)}\,,\label{fermi-nucl}\end{equation}
 where the parameter $c$ is the half charge radius and $a$ is related
to skin thickness, defined as the interval of the nuclear thickness
in which the nuclear charge density falls from near one to near zero.
We have taken a large basis set to check the convergence of the results
on the number of basis functions used. Excitations from all the core
electrons have been considered for all the cases. The details of the
basis sets used in the calculations presented here are given in table
\ref{basis}.

\begin{table}

\caption{\label{CII-results}Ionization potential (IP) and the excitation
energies (EEs) (in $\mathrm{cm^{-1}})$ for C II. The column `Koopman'
contains the Dirac-Fock energies and the columns designated as (T)
contain the effects of partial triple excitations in the valence sector.
Observed values of IP and EEs are taken from the NIST table \cite{NIST}
unless mentioned otherwise.}

~

\begin{centering}
\begin{tabular}{rllllll}
\hline 
State&
Koopman &
CCSD&
CCSD(T)&
UCCSD&
UCCSD(T)&
Observed\tabularnewline
\hline
\hline 
IP~~ $2p\,^{2}P_{1/2}$&
189794.81&
196575.36&
197825.00&
196739.57&
197988.30&
196592.44 \cite{moore}\tabularnewline
EE~~$2p\,^{2}P_{3/2}$&
73.28&
73.44&
-17.22&
45.37&
44.67&
63.42\tabularnewline
$3s\,^{2}S_{1/2}$&
110674.88&
109729.32&
108025.07&
108203.18&
105768.50&
116537.65\tabularnewline
$3p\,^{2}P_{1/2}$&
127422.46&
131623.85&
132703.83&
131766.92&
132838.66&
131724.37\tabularnewline
$3p\,^{2}P_{3/2}$&
127433.78&
131636.80&
132726.16&
131780.30&
132860.20&
131735.52\tabularnewline
\hline
\hline 
&
&
&
&
&
&
\tabularnewline
\end{tabular}
\par\end{centering}
\end{table}

\begin{table}

\caption{\label{RbI-results}Ionization potential (IP) and the excitation
energies (EEs) (in $\mathrm{cm^{-1}})$ for Rb I. Observed values
given in the last column are taken from the NIST table \cite{NIST}.}

~

\begin{centering}
\begin{tabular}{rcccccc}
\hline 
State&
Koopman&
CCSD&
CCSD(T)&
UCCSD&
UCCSD(T)&
Observed\tabularnewline
\hline
\hline 
IP~~ $5s\,^{2}S_{1/2}$&
30592.05&
33690.23&
33694.39&
33691.16&
33694.91&
33690.57\tabularnewline
EE~~$5p\,^{2}P_{1/2}$&
10660.57&
12610.94&
12594.57&
12611.16&
12594.81&
12578.95\tabularnewline
$5p\,^{2}P_{3/2}$&
10898.41&
12850.35&
12849.30&
12850.59&
12849.54&
12816.54\tabularnewline
$4d\,^{2}D_{5/2}$&
17494.13&
19484.27&
19444.28&
19483.57&
19430.27&
19355.20\tabularnewline
$4d\,^{2}D_{3/2}$&
17481.27&
19482.91&
19434.31&
19482.19&
19435.20&
19355.65\tabularnewline
\hline
\hline 
&
&
&
&
&
&
\tabularnewline
\end{tabular}
\par\end{centering}
\end{table}

\section{\label{analysis}Analysis and discussions}

Tables \ref{CII-results} and \ref{RbI-results} present the ionization
potential (IP) for the ground state and the excitation energies (EEs)
for the few low lying excited states for C II and Rb I, respectively.
From the calculations and the tabulated results we have observed a
nice feature about the usage of perturbative triple excitations often
used in the CC calculations. From Table \ref{CII-results} we have
observed that for singly ionized C I \emph{i.e.} C II (an element
in the group IV in the periodic table), CCSD method works reasonably
well to estimate the IP of the ground state and EE for the first excited
state, whereas the UCCSD method performs better for estimating the
EEs of the excited states like $3p\,^{2}P_{1/2}$ and $3p\,^{2}P_{3/2}$.
When we consider the effect of perturbative partial triple excitations
for the valence electron, namely CCSD(T), the method fails miserably
to estimate the IP and the EE's. Moreover CCSD(T) even fails to determine
the ground state of C II. This is reflected in the value of the EE
of the $2p\,^{2}P_{3/2}$ state which has a negative sign. That indicates
CCSD(T) determines $2p\,^{2}P_{3/2}$ to be the ground state of C
I instead of $2p\,^{2}P_{1/2}$. On the other hand, when we apply
UCCSD(T) to estimate IP and EEs for C II it performs better than CCSD(T)
but still is not good enough to calculate them accurately as compared
to CCSD and UCCSD. Moreover UCCSD(T) is also capable of determining
$2p\,^{2}P_{1/2}$ as the ground state of C II.

Table \ref{RbI-results} contains the IP of the ground state and EEs
for the first few excited states of the alkali atom Rb which is positioned
in the Gr-I in the periodic table. We have observed that both CCSD/UCCSD
perform better to determine the IP of the ground state and the EE
for the first excited states. Whereas, to determine the EEs of the
second excited state and onwards, the partial triple excitations from
the valence sector contribute quite significantly. If we do a close
comparison to the effects of partial triple excitations in the CCSD
and UCCSD level, denoted by CCSD(T) and UCCSD(T) respectively we can
find out that UCCSD(T) even performs better to determine the EE's
of the high lying excited states.

\begin{figure}
\begin{centering}
\includegraphics{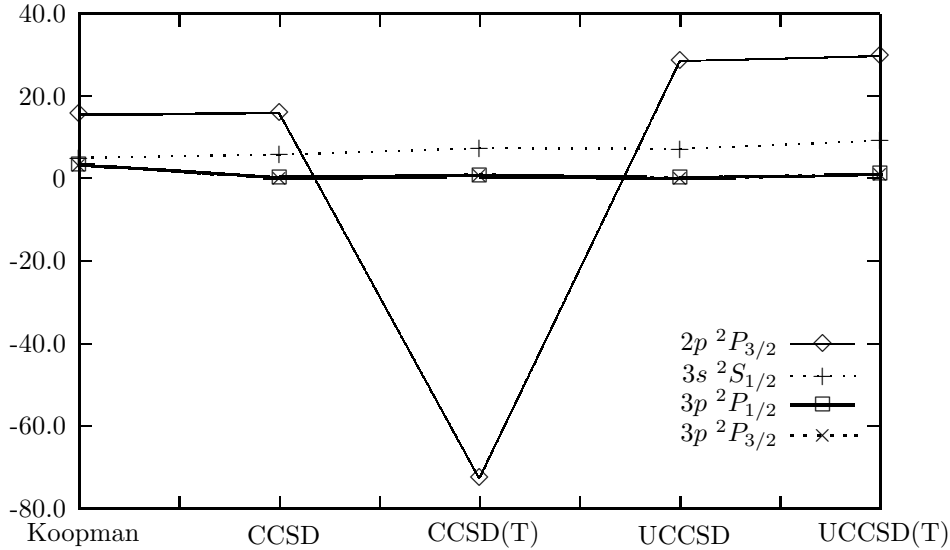} 
\par\end{centering}

\caption{\label{CII-error}Relative error (in $\%$) in estimation of EEs
for different states of C II. The acronyms for the different methods
are discussed in the text and in the tables.}
\end{figure}

\begin{figure}
\begin{centering}
\includegraphics{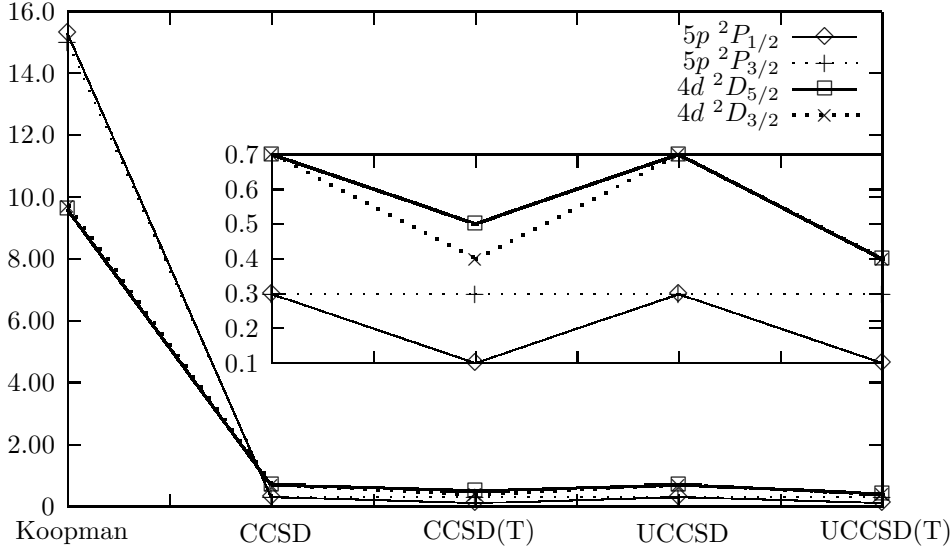} 
\par\end{centering}

\caption{\label{RbI-error}Relative error (in $\%$) in estimation of EEs
for different states of Rb I. The figure shown in the inside box is
a magnification of the relative errors for different CC methods. }
\end{figure}

We have shown earlier \cite{csur-ucc} that unlike CCSD, UCCSD can
contains more effects from higher order excitations in the same level
of excitation because of the structure of the core excitation operator.
Consideration of higher order excitation is the key point to understand
the improved performance of UCCSD for the high lying excited states.
This has been discussed in section \ref{unitary}. In CCSD(T) and
UCCSD(T) we have considered the effect of partial triples denoted
by (T) in the valence sector. If we take a close look at the electronic
structure of the atoms considered in the calculations we will find
that for C II, the core is defined as $1s^{2}2s^{2}$ and the L-shell
(with principal quantum number 2) is not completely filled. On the
other hand for Rb I the core is defined as $[\mathrm{Kr}]$ which
has a completely filled N-shell (principal quantum number 4). When
we apply CCSD(T) or UCCSD(T) for C II, because of the unfilled L-shell,
the correlation effect, the most important many-body effects in multi-electron
atoms, between the valence electron and the electrons from the unfilled
L-shell (in this case the $2s$ electrons) turns out to be very important.
This is reflected in the Koopman energies listed in the table. More
explicitly, although the $2s$ sub-shell ($\mathrm{L_{1}}$ shell)
is fully occupied in the ground state of C II, but the vacancy in
the rest of the L shell makes the case a little different than Rb
I. In figure \ref{CII-error} and \ref{RbI-error} we have graphically
shown our findings. In fig \ref{RbI-error} the figure given in the
inside box contains the relative errors (in $\%$) in estimating the
EEs of first few excited states using different CC methods. This inner
figure helps us to see the relative errors for the different CC methods
in a proper scaling.

To generalize our findings we have also studied two more systems,
Li I, the alkali atom with filled K-shell but unfilled L-shell (Ground
state of Li I : $1s^{2}2s$) and Al I with the ground state configurations
$1s^{2}2s^{2}2p^{6}3s^{2}3p$ (unfilled M shell). Both Li I and Al
I have similarities with Rb I and C II respectively in terms of the
vacancies in the principal shell. Earlier we have reported the determination
of the IP and EEs for Al I using CCSD and UCCSD \cite{csur-al}. For
the alkali atoms like Li I and Rb I, the valence electrons feel the
potential of a core with a completely filled principal shell (K and
M shell respectively). On the other hand for C II and Al I the core
do not have a completely filled principal shell. For these two atoms
the electron correlation between the electrons in the unfilled principal
shell play important roles in determining the state energies. In this
study we have found the similar pattern of performance of CCSD/UCCSD
and CCSD(T)/UCCSD(T) for systems with filled/unfilled principal shell
for the core state.

\section{\label{conclusion}Conclusion}

In conclusion, we want to focus on the findings of our work in the
following way. The contribution of partial triples through CCSD(T)/UCCSD(T)
method works well for atoms or ions with a filled principal shell
in the core. On the other hand if there is a vacancy in the principal
shell in the core the coupled cluster (CC) and the unitary coupled
cluster (UCC) method with partial triple excitations in the valence
sector fails miserably. In general CCSD method works well to determine
the IP of the ground state and the EE of the first excited state.
Whereas, the unitary counterpart of CCSD, namely UCCSD performs better
to determine the EEs for the high lying excited states.

To our knowledge this is the first attempt to analyze the effects
of partial triple excitation in atomic coupled cluster calculations
in this manner. One can generalize our findings to estimate the state
energies for atoms/ions with an unfilled principal shell in their
configuration. The present study clearly demonstrates that CCSD(T)
is, in general, not the best method for accurate determination of
state energies for atoms with a single valence electron. This is important
because CCSD(T) is used to estimate the error in the theoretical determination
of state energies and atomic properties like transition probabilities
and expectation values \cite{csur-prl}. Our findings in this work
will focus on the issue to search for a new method for estimating
the error.

\begin{verse}
\textbf{Acknowledgment} : These computations are carried out in the
Intel Xeon cluster at the Department of Astronomy, OSU under the Cluster-Ohio
initiative. This work was partially supported by the National Science
Foundation and the Ohio State University (CS). One of the author (RKC)
thanks Prof. Karl F Freed, University of Chicago, for hospitality.
RKC acknowledges the Department of Science and Technology, India (grant
SR/S1/PC-32/2005). 
\end{verse}

\end{document}